\documentclass[aps,prl,twocolumn,groupedaddress,showpacs,preprintnumbers,
amsmath,amssymb]{revtex4}
\usepackage[dvips]{graphicx}
\usepackage{amsfonts}
\usepackage{amssymb}
\usepackage{float}
\usepackage{natbib}
\usepackage{amsmath}
\usepackage{float}
\usepackage{dcolumn}
\usepackage{lscape}

\renewcommand{\d}{{\rm d}}

\newcommand{\ggt}{\widetilde{g}}

\newcommand{\lambdab}{\widetilde{\lambda}}

\makeatletter
\newlength{\earraycolsep}
\def\eqnarray{\stepcounter{equation}\let\@currentlabel%
\theequation
\global\@eqnswtrue\m@th
\global\@eqcnt\z@\tabskip\@centering\let\\\@eqncr
$$\halign to\displaywidth\bgroup\@eqnsel\hskip\@centering
$\displaystyle\tabskip\z@{##}$&\global\@eqcnt\@ne
\hskip 2\earraycolsep \hfil$\displaystyle{##}$\hfil
&\global\@eqcnt\tw@ \hskip 2\earraycolsep
$\displaystyle\tabskip\z@{##}$\hfil
\tabskip\@centering&\llap{##}\tabskip\z@\cr}
\makeatother

\begin{document}

\title{Closed-Orbit Theory of Spatial Density Oscillations in Finite
  Fermion Systems}

\author{J\'er\^ome Roccia and Matthias Brack}

\affiliation{Institut f\"ur Theoretische Physik, Universit\"at Regensburg,
             D-93040 Regensburg, Germany}

\date{\today}

\begin{abstract} 
We investigate the particle and kinetic-energy densities for $N$ 
non-interacting fermions confined in a local potential. Using  
Gutzwiller's semi-classical Green function, we describe the oscillating 
parts of the densities in terms of closed non-periodic classical orbits. 
We derive universal relations between the oscillating parts of the 
densities for potentials with spherical symmetry in arbitrary 
dimensions, and a ``local virial theorem'' valid also for arbitrary 
non-integrable potentials. We give simple analytical formulae for the 
density oscillations in a one-dimensional potential.
\end{abstract}



\maketitle

{\it Introduction.--- }
Finite systems of fermions are studied in many branches of physics, 
e.g., electrons in atoms, molecules, and quantum dots; protons and 
neutrons in atomic nuclei; or fermionic atoms in traps. Common to 
these systems are pronounced shell effects which result from the 
combination of quantized energy spectra with the Pauli exclusion 
principle. The shell effects manifest themselves most clearly in 
ionization (or separation) energies and total binding energies. 
They lead to ``magic numbers'' of particles in particularly stable 
species, when degenerate shells or approximately degenerate bunches 
of single-particle levels are filled. Shell effects appear also in 
spatial particle densities \cite{ks,thto} and kinetic-energy densities. 
Near the center of a system, the alternating parities of the occupied 
shells lead to regular quantum oscillations, while the so-called 
``Friedel oscillations'' characteristically appear near the surface 
of a sufficiently steep confining potential. Kohn and Sham \cite{ks} 
analyzed both oscillations in the particle density using Green 
functions in the one-dimensional WKB approximation. Thouless and 
Thorpe \cite{thto} extended their method to give analytical results 
also for the central oscillations in three-dimensional systems with 
radial symmetry.

In the periodic orbit theory (POT) \cite{gutz,babl,poch},  semi-classical 
``trace formulae'' allow one to relate the level density of a quantized 
Hamiltonian system to the periodic orbits of the corresponding classical 
system. This can be used to interpret quantum shell effects occurring 
in finite fermion systems in terms of the shortest periodic orbits 
(see Ref.\ \cite{book} for an introduction to POT and applications to 
various branches of physics). To our knowledge, no attempt has been 
made so far to interpret quantum oscillations of spatial densities in 
terms of classical orbits. In the present paper, we use the 
semi-classical Green function of Gutzwiller \cite{gutz} to derive 
analytical expressions for the oscillating parts of particle and
kinetic-energy densities in terms of closed non-periodic classical 
orbits.

{\it General framework.--- }
We consider a $D$-dimensional system of $N$ non-interacting particles 
with mass $m$, which obey Fermi-Dirac statistics and are bound by a 
local potential $V({\bf r})$. Note that $V({\bf r})$ can be the 
self-consistent mean field of an {\it interacting} fermion system 
(such as a nucleus). The energy eigenvalues $E_n$ and 
eigenfunctions $\psi_n({\bf r})$ are given by the stationary
Schr\"odinger equation. The particle density of the system at zero 
temperature, ignoring the spin degeneracy, is given by
\begin{equation} \label{rho}
\rho({\bf r})= \sum_{E_n \leq \lambda(N)} \psi_n^{\star}({\bf r}) \psi_n({\bf r})\,,
\end{equation}
where the Fermi energy $\lambda(N)$ is determined by normalizing the 
density to the given particle number $N$.
For the kinetic-energy density we discuss two different forms
\begin{eqnarray}
\tau({\bf r})&=&- \frac{\hbar^2}{2m}\sum_{E_n \leq \lambda}
 \psi_n^{\star}({\bf r}) \nabla^2\psi_n({\bf r}) \,,   \label{tau}\\
\tau_1({\bf r})&=& \frac{\hbar^2}{2m}\sum_{E_n \leq \lambda}
 |\nabla \psi_n({\bf r})|^2  \,, \label{tau1}
\end{eqnarray}
which after integration both lead to the exact total kin\-etic energy. 
We rewrite the above densities in the form
\begin{eqnarray}
\rho({\bf r})&=&-\frac{1}{\pi}\, \Im m \int_0^{\lambda} \d E \, G(E,{\bf r},{\bf r}')
|_{{\bf r'}={\bf r}} 
\,,  \label{rhog} \\
\tau({\bf r})&=& \frac{\hbar^2}{2\pi m}\, \Im m \int_0^{\lambda}  \d E \, 
\nabla^2_{{\bf r}'}G(E,{\bf r},{\bf r}') |_{{\bf r}'={\bf r}} 
 \,, \label{taug}\\
\tau_1({\bf r})&=&
-\frac{\hbar^2}{2\pi m}\, \Im m \int_0^{\lambda} \d E  \,
\nabla_{{\bf r}}\nabla_{{\bf r}'}G(E,{\bf r},{\bf r}')  |_{{\bf r}'={\bf r}}
 \,, \ \ \ \ \label{tau1g}
\end{eqnarray}
where $G(E,{\bf r},{\bf r}')$ is the Green function in the energy representation
\begin{equation}
G(E,{\bf r},{\bf r}')= \sum_n \frac{\psi_n^{\star}({\bf r}) 
                       \psi_n({\bf r}')}{E+i \epsilon -E_n} \,, 
                       \quad (\epsilon >0) \,,
\end{equation}
and the identity $\displaystyle 1/(E+i \epsilon -E_n)={\cal P}
[1/(E-E_n)]-i \pi \delta (E-E_n)$ is used (${\cal P}$ is the Cauchy 
principal value).

To obtain semi-classical expressions, we replace the Green 
function by Gutzwiller's approximation \cite{gutz}
\begin{equation}\label{Green}
G_{scl}(E,{\bf r,r'})=\alpha_{_D}\!\!\sum_{\rm cl. trj.}\!\!\!
\sqrt{|{\cal D}|}
\,e^{\frac{i}{\hbar} S(E,{\bf r,r'})-i \mu \frac{\pi}{2}},
\end{equation}
which is valid to leading order in $1/\hbar$ in the semi-classical
limit $\hbar\to0$, i.e., when the dominating classical actions
$S(E,{\bf r,r'})$ are large compared with $\hbar$. In Eq.~(\ref{Green}),
${\cal D}$ is the Van Vleck determinant given below, $\mu$ is the 
Morse index and $\alpha_{_D}=2 \pi (2 i \pi \hbar)^{-(D+1)/2}$. The sum 
is over all classical trajectories starting at ${\bf r}$ and ending 
at ${\bf r'}$. The action integral along each trajectory is
\begin{equation}
S(E,{\bf r,r'})=\int_{\bf r}^{\bf r'}\!\! {\bf p}({\bf r''})\cdot \d {\bf r''}.
\label{action}
\end{equation}
Since we have to use ${\bf r}={\bf r'}$ in 
(\ref{rhog}) - (\ref{tau1g}), only closed trajectories 
starting and ending at the same point ${\bf r}$ have to be included in 
the sum of (\ref{Green}). Following Gutzwiller \cite{gutz},
we use for each trajectory a local coordinate system
${\bf r}=(q,{\bf r}_{\bot})=(q,r_{\bot1},r_{\bot2},\dots,r_{\bot(D-1)})$,
whose first variable $q$ is chosen along the trajectory, while the 
vector ${\bf r}_{\bot}$ of the remaining $D-1$ variables is transverse 
to it. The Van Vleck determinant then becomes
\begin{equation}\label{eq_vv_det2}
{\cal D}=\frac{(-1)^D\,m^2\,{\cal D}_{\bot}}{p(E,{\bf r})\,p(E,{\bf r'})}, \qquad
{\cal D}_{\bot}=\det (\partial{\bf p'}_{\!\!\bot}/\partial{\bf r}_{\!\bot})\,,
\end{equation}
where ${\bf p}(E,{\bf r})=\dot{{\bf r}}\left|\!\sqrt{2m[E-V({\bf
r})]}\right|\!/|{\dot{\bf r}}|$ is the classical momentum and 
$p(E,{\bf r})=m\,\dot{q}(E,{\bf r})$ its modulus.

We now want to keep only the leading-order terms in the semi-classical
expansion parameter $\hbar$. To this purpose it is useful to decompose
the Fermi energy into a smooth and an oscillating part: $\lambda=
\lambdab + \delta\lambda$. Assuming that $\delta \lambda \ll \lambdab$, one
can show \cite{clmr} that
\begin{equation}\label{eq_form_lambt}
\delta \lambda=-\int_0^{\lambdab}\! \delta g(E)\,\d E\,/\,\ggt(\lambdab) \,,
               \quad \int_0^{\lambdab}\ggt(E)\,\d E=N\,,
\end{equation}
where $\ggt(E)$ is the smooth part of the level density and 
$\delta g(E)$ its oscillating part, semi-classically given by a sum 
over the periodic orbits of the classical system \cite{gutz}. 

The sum over closed trajectories to be used in (\ref{rhog}) - (\ref{tau1g}) 
can be separated into a sum over periodic orbits (POs) and a sum over 
non-periodic orbits (NPOs). The actions along the POs are independent 
of ${\bf r}$; their contributions are therefore smooth functions, given 
only by the initial and final momenta ${\bf p}={\bf p}(E,{\bf r})$ and 
${\bf p'}={\bf p}(E,{\bf r'})$. To lowest order in $\hbar$, the 
semi-classical densities are given by the POs {\it with zero length}.
They are identical with the smooth Thomas-Fermi (TF) densities \cite{marc}, 
like it is known \cite{bm} for the level density $\ggt(E)=g_{TF}(E)$. To 
next order in $\hbar$, the sums over all NPOs yield the density
oscillations, so that the semi-classical particle density has the form
\begin{equation}
\rho_{scl}({\bf r}) = \rho_{TF}({\bf r})+\delta\rho({\bf r})\,.
\label{rhosc}
\end{equation}
Analogous forms hold for $\tau({\bf r})$ and $\tau_1({\bf r})$

For the kinetic-energy densities we have to derive the Green function
(\ref{Green}) twice according to (\ref{taug},\ref{tau1g}). The 
semi-classically leading terms come from the derivatives of 
$S(E,{\bf r,r'})$, for which the relations
$\nabla_{{\bf r'}} S(E,{\bf r,r'})={\bf p'}$ and $\nabla_{{\bf r}} 
S(E,{\bf r,r'})=-{\bf p}$ hold. The energy integration in Eqs.\ 
(\ref{rhog}) - (\ref{tau1g}) can be done by parts. The leading-order
results come from the upper integration limit, taken as $\lambdab$. 
The lower limit, which must be taken to be $V({\bf r})$ since in 
the semi-classical approximation one has to stay in the classically 
allowed region, gives no contributions.
We then obtain for the oscillating parts of the densities:
\begin{eqnarray}
\delta \rho({\bf r})&=&\frac{m\hbar}{\pi}\, \Re e \ \alpha_{_D}\!\sum_{\rm NPO}
   \frac{\sqrt{|{\cal D}_{\bot}|}_{{\bf r'}={\bf r}}}{p(\lambdab,{\bf r})\, T(\lambdab,{\bf r})}
   \, e^{i\Phi(\lambdab,{\bf r})},
\label{drhosc}\\
\delta \tau({\bf r})&=& \frac{\hbar}{2\pi}\, \Re e \ \alpha_{_D}\!\sum_{\rm NPO}
   \frac{p(\lambdab,{\bf r})\sqrt{|{\cal D}_{\bot}|}_{{\bf r}'={\bf r}}}{T(\lambdab,{\bf r})}\, 
   e^{i\Phi(\lambdab,{\bf r})}, 
\label{dtausc}\\
\delta \tau_1({\bf r})&=& \frac{\hbar }{2\pi}\, \Re e \ \alpha_{_D}\!\!\sum_{\rm{NPO}}\!
   \frac{\!\{({\bf p}\!\cdot\!{\bf p'})_{\lambda}\!\sqrt{|{\cal D}_{\bot}|}\}_{{\bf r}'={\bf r}}}
   {p(\lambdab,{\bf r})\, T(\lambdab,{\bf r})}\, e^{i\Phi(\lambdab,{\bf r})}\!,~~~~
\label{dtau1sc}
\end{eqnarray}
where $({\bf p}\cdot{\bf p'})_\lambda={\bf p}(\lambdab,{\bf r})
\cdot{\bf p}(\lambdab,{\bf r'})$, the phase function in the 
exponents is $\Phi(\lambdab,{\bf r})=S(\lambdab,{\bf r,r})/\hbar
-\mu\frac{\pi}{2}$, and $T(\lambdab,{\bf r})=\d S(E,{\bf 
r,r})/\d E|_{E=\lambdab}$. Since the modulus $p$ depends only on 
position and Fermi energy, but not on the orbits, we can take it 
outside the sum over the NPOs. We thus immediately find the general 
relation
\begin{equation}\label{taurho}
\delta \tau({\bf r})=[\lambdab-V({\bf r})] \,\delta \rho({\bf r}) \,.
\end{equation}
It holds for arbitrary, integrable or non-integrable, local potentials 
in arbitrary dimensions. Eq.\ (\ref{taurho}) may be termed 
a ``local virial theorem'' because it relates kinetic and potential 
energy densities locally at any point.
For $\delta\tau_1({\bf r})$ we have no such relation, since it
depends on the relative directions of final and initial momentum
of each orbit. Due to the semi-classical nature of our approximation,
Eq.~(\ref{taurho}) and the results derived below are expected
to be valid in the limit of large particle numbers $N$.

{\it One-dimensional systems.--- }
For the further development we now focus on one-dimensional systems 
characterized by a smooth binding potential $V(x)$ with a minimum at $x=0$. 
We will explicitly derive a semi-classical expression for the particle 
density $\rho(x)$; analogous results for the kinetic densities are 
found in the same way.

The classical motion at fixed energy $E$ is limited by the turning 
points $x_\pm(E)$ defined by $V(x_\pm)=E$, with $x_+(E)>0$ and
$x_-(E)<0$. In one dimension there are only two types of trajectories 
going from $x$ to $x'$: the first type has its momenta at the initial 
and final points in the same direction, while for the second type they 
go in opposite directions. Without loss of generality we may choose 
$x_-\leq x\leq x'\leq x_+$. The shortest trajectory of the first type 
goes from $x$ directly to $x'$ without reaching any of the turning
points; it is indexed by the subscript '0' and has the action
\begin{equation}
S_{0}(E,x,x') = S(E,0,x')-S(E,0,x) \,.
\label{eq_ac_tf}
\end{equation}
All other trajectories of the first type bounce $j=1,2,\dots$ times
forth and back between the turning points before reaching $x'$; they
are indexed by '1' and have the actions
\begin{equation}
S_{1\pm}(E,x,x') = j\,S_1(E) \pm S_0(E,x,x')\,, \quad (j=1,2,\dots) 
\label{eq_ac_po}
\end{equation}
where $S_1(E)$ is the action of the primitive periodic orbit and the 
sign $\pm$ refers to the starting direction. The trajectories of the 
second type bounce $k=0,1,\dots$ times forth and back before reaching
$x'$; they are indexed by '2' and have the actions 
\begin{eqnarray}
S_{2\pm} (E,x,x') & = & k S_1(E) + 2 S_{\mp}(E)\!\!\!\!\!\!
                        \label{eq_ac_npo+}\\
                 & \pm & [S(E,0,x')+S(E,0,x)]\,,\;\; (k=0,1,\dots\!)
                   \nonumber 
\end{eqnarray}
where
$S_-(E)=S(E,x_-,0)$ and $S_+(E)=S(E,0,x_+)$. For a symmetric potential
with $V(x)=V(-x)$, one has $S_-(E)=S_+(E)=S_1(E)/2$. From 
(\ref{eq_vv_det2}) we have $\sqrt{|{\cal D}(E,x,x)|}=m/p(E,x)$
for all trajectories. For smooth potentials in one dimension, the
Morse index $\mu$ is equal to the number of turning points, which for 
the above trajectories is $\mu_0 =0$, $\mu_{1\pm}=2j$, and $\mu_{2\pm}=2 k+1$. 
[For a one-dimensional box with reflecting walls, the Morse index 
equals twice the number of turning points; our semi-classical densities 
become exact in this case.] 

Using (\ref{eq_ac_tf}--\ref{eq_ac_npo+}) and $D$=1 in Eq.\ 
(\ref{Green}), we now obtain the semi-classical particle density 
$\rho_{scl}(x)$ as a sum of the three types of contributions indexed 
as above:
\begin{equation}\label{eq_sum_rho}
\rho_{scl}(x)  = \rho_0(x)+\!\sum_{\sigma=+-}\!\!
                 [\,\rho_{1\sigma}(x)+\rho_{2\sigma}(x)\,]\,.
\end{equation}
Since we have to use $x=x'$, the only contributing orbits of type 0
have zero length, those of type 1 are periodic, and those of type
2 are non-periodic. Doing the energy integration by parts, we get
to leading-order in $\hbar$ 
\begin{eqnarray}
\rho_0(\lambda,x)&=&\frac{(2m)^{1/2}}{\pi\hbar}\sqrt{\lambda-V(x)}\,,
\label{eq_rhob}\\
\rho_{1}(x)&=&\frac{2m}{\pi} \sum_{j=1}^{\infty} (-1)^j\,
\frac{\sin\{jS_1(\lambdab)/\hbar\}}{p(\lambdab,x)\,jT_1(\lambdab)}\,,
\label{eq_rho_1}
\end{eqnarray} 
where $T_1(\lambdab)$ is the period of the primitive periodic orbit.
Taylor expanding $\rho_0(\lambda,x)$ in Eq.\ (\ref{eq_rhob}) around 
$\lambdab$ yields the 
well-known TF density, $\rho_{TF}(x) = \rho_0(\lambdab,x)$,
plus a term linear in $\delta\lambda$ which, using Eq.\ 
(\ref{eq_form_lambt}), cancels exactly the contribution 
$\rho_1(x)$ in (\ref{eq_rho_1}). The leading-order oscillating 
term is therefore given by the type 2 orbits, i.e., by 
$\delta\rho(x)=\rho_{2+}(x)+\rho_{2-}(x)$ which has the explicit form
\begin{equation}
\delta\rho(x) = -\frac{m}{\pi}\! \sum_{k=0\atop\sigma=\pm}^\infty (-1)^k
                 \frac{\cos\{[k S_1(\lambdab)\!+\!R_{\sigma}(\lambdab,x)]/\hbar\}}
                 {p(\lambdab,x)[kT_1(\lambdab)+R'_{\sigma}(\lambdab,x)]},
\label{delrhogen}
\end{equation} 
with $R_{\pm}(\lambdab,x)=2S_{\pm}(\lambdab)\mp2S(\lambdab,0,x)$.
This result is equivalent, although not obviously identical,
with the result given in Eq.\ (3.36) of \cite{ks}.

For the kinetic-energy densities we proceed in the same way. The
smooth parts of $\tau(x)$ and $\tau_1(x)$ are identical and
equal to the TF kinetic-energy density $\tau_{TF}(x)$; for their 
oscillating parts we obtain the one-dimensional version of the 
relation (\ref{taurho}) and, in addition, the new relation
\begin{equation}\label{tautau}
\delta \tau_1(x)=-\delta\tau(x) \,,
\end{equation}
which holds due to the opposite initial and final momenta of the NPOs 
of type 2 which contribute to (\ref{dtau1sc}).

\begin{figure}[ht]
\includegraphics[width=0.8\columnwidth,clip=true]{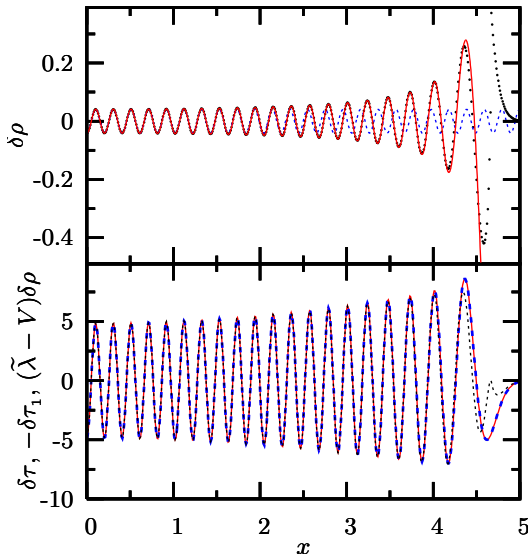} 
\caption{\label{fig1}
(Color online)
{\it Upper panel:}
Oscillating part $\delta\rho(x)$ of the particle density of 
$N$=40 fermions in the potential $V(x)=x^4\!/4$. Dots show the 
quantum-mechanical result; the solid line shows the semi-classical 
result $\delta\rho(x)$ in (\ref{delrhogen}) and the dashed line
the approximation (\ref{delrho}) for small $x$ values.\\
{\it Lower panel:}
Oscillating parts of the quantum-mechanical kinetic-energy 
densities in the same system:
$\delta\tau(x)$ (solid line) and $-\delta\tau_1(x)$ (dashed line).
The dotted line shows the function $[\lambdab-V(x)]\,\delta\rho(x)$
using the quantum-mechanical $\delta\rho(x)$.}
\end{figure}

In Fig.\ \ref{fig1}, we test our semi-classical results for the potential 
$V(x)=x^4\!/4$ with $N=40$ particles (with units such that
$\hbar=m=1$). The upper panel shows $\delta\rho(x)$ given in
(\ref{delrhogen}) by the solid line, while the dots represent 
the quantum-mechanical expression (\ref{rho}) after subtracting 
the TF density. The agreement is very good except close to  
the classical turning point where the TF approximation 
breaks down. The lower panel demonstrates the validity of the 
relations (\ref{taurho}) (with ${\bf r}\to x$) and (\ref{tautau}). 
The small deficiencies near the classical turning points can be
overcome and the tail in the classically forbidden region described
by the standard WKB treatment \cite{ks,thto} or the TF-Weizs\"acker 
theory \cite{marc}. 

A simpler form for $\delta\rho(x)$ is found if one restricts oneself
to the interior part of the system around $x=0$, where $V(x)\ll\lambdab$. 
Then the action integral $S(\lambdab,0,x)$ can be approximated by 
$S(\lambdab,0,x)\simeq xp_\lambda$, where $p_\lambda=(2m\lambdab)^{1/2}$ 
is the smooth Fermi momentum. We then obtain
\begin{equation} \label{eq_rho_2+}
\delta\rho(x) = \frac{-2 m\cos (2xp_\lambda/\hbar+\delta\Phi)}{\pi\,p_\lambda\,
                T_1(\lambdab)}\,C_N(\lambdab)\,, 
\end{equation}
where $\delta\Phi= [S_-(\lambdab)-S_+(\lambdab)]/\hbar$ is a phase difference 
related to the asymmetry of the potential, and
\begin{equation}
C_N(\lambdab) = \sum_{k=0}^{\infty} (-1)^k 
      \frac{\cos\lbrace[(k+1/2) S_1(\lambdab)]/\hbar\rbrace}{(k+1/2)}\,.
\label{orbitsum}
\end{equation}
To evaluate this sum, we exploit the fact that the action $S_1(\lambdab)$
in one dimension can be related to the particle number
$N$ by $S_1(\lambdab)\approx 2\pi\hbar N$, which is nothing but the 
well-known Bohr-Sommerfeld quantization condition. Using this 
relation in (\ref{orbitsum}) and the identity
{\small $\sum_{k=0}^{\infty} 
(-1)^k \cos[(2k+1)N]/(2k+1)=(-1)^N\pi/4$}, we find the approximate
expression for the central oscillations
\begin{equation}
\delta \rho(x) = (-1)^{N+1}\,\frac{m}{p_\lambda T_1(\lambdab)}\,
                             \cos(2xp_\lambda/\hbar+\delta\Phi)\,,
\label{delrho}
\end{equation}
which can also be obtained from Eq.\ (3.36) in Ref.\ \cite{ks} 
in the limit $V(x)\ll\lambdab$. It is shown by the dashed line in the
upper panel of Fig.\ \ref{fig1}.
Using Eqs.\ (\ref{taurho}) and (\ref{tautau}) one can give analogous
simple results for the kinetic-energy density oscillations near $x=0$.

Our derivation shows that {\it periodic orbits} do {\it not} contribute to the 
oscillations in the densities $\rho(x)$, $\tau(x)$ and $\tau_1(x)$, 
while they are known \cite{gutz} to give the most important contributions 
to the oscillating level density $\delta g(E)$. In fact, 
the most important contribution to (\ref{delrhogen}) comes from the two 
{\it shortest non-periodic} orbits which go from $x$ 
to one of the turning points and back; for small $x$ their action 
difference is $2xp_\lambda$. The summation over all longer non-periodic
orbits yields the oscillating sign depending 
\noindent
on the particle number 
$N$. We emphasize that the oscillations in Eq.\ (\ref{delrho}) have
the universal wave length $\hbar\pi/p_\lambda$ independent on the
particular form of the potential $V(x)$.

\newpage

{\it Higher-dimensional radial systems.--- }
In a forthcoming paper \cite{bkmr}, we generalize our method for
higher-dimensional systems. For binding potentials $V(r)$ with 
spher\-ical symmetry in $D>1$ dimensions, one can separate
two kinds of spatial oscillations in the radial variable:\\ 
$(i)$ irregular longer-ranged oscillations, which are 
attrib\-uted to nonlinear classical orbits, and\\ 
$(ii)$ regular, rapid oscillations of the kind discussed above
and denoted here by $\delta\rho(r)$, $\delta\tau(r)$, and
$\delta\tau_1(r)$.\\
The regular 
rapid oscillations originate from non-periodic {\it linear} 
orbits with zero angular momentum, start\-ing from $r$ in the radial 
direction and returning with opposite radial momentum to $r$; these 
orbits correspond exactly to our above type 2 orbits in one dimension.
From their contributions to the semi-classical Green function 
(\ref{Green}) and hence to (\ref{drhosc}), it is straightforward to 
derive the following relation, valid to leading order in $\hbar$:
\begin{equation}
-\frac{\hbar^2}{8m}\nabla^2\delta\rho(r) = [\lambdab-V(r)]\,\delta\rho(r)\,.
\label{Laplace}
\end{equation}
Similarly, it follows from the nature of the radial type 2 orbits 
that $({\bf p}\cdot{\bf p'})_{{\bf r}'={\bf r}}=-p^2$ in (\ref{dtau1sc}) 
and hence the rapid oscillations in the kinetic-energy densities 
$\tau(r)$ and $\tau_1(r)$ fulfill the relation (\ref{tautau}) 
in the radial variable $r$:
\begin{equation}
\delta\tau_1(r) = - \delta\tau(r)\,.
\label{tautaurad}\vspace*{0.1cm}
\end{equation}
For small $r$, where $V(r)\ll\lambdab$, Eq.\ (\ref{Laplace})
becomes a universal eigenvalue equation for $\delta\rho(r)$ 
with eigenvalue $\lambdab$, which can be transformed into the 
Bessel equation. Its solutions yield the generalization of 
Eq.\ (\ref{delrho}) (with $\delta\Phi=0$) for the rapid oscillations
near $r=0$:
\begin{equation}
\delta \rho(r) = (-1)^{^{M\!-1}}\frac{m}{2\hbar\,T_{r1}(\lambdab)}
                 \left(\frac{p_\lambda}{4\pi\hbar r}\right)^{\!\nu}
                 \!\!J_\nu(2rp_\lambda/\hbar)\,.
\label{delrhorad}
\end{equation}
Here $J_\nu(z)$ is a Bessel function with index $\nu=D/2-1$,
$M$ is the number of filled main shells, and 
$T_{r1}$ is the period of one radial oscillation.
For $D=3$, Eq.~(\ref{delrhorad}) agrees with the result of 
\cite{thto} up to a $\lambdab$ dependent normalization factor.
Our results (\ref{taurho}) and (\ref{Laplace}) - (\ref{delrhorad}) 
agree with those derived analytically for harmonic oscillator
potentials $V(r)=cr^2$ in arbitrary dimension $D$ from the 
{\it quantum-mechanical} densities to leading order in a 
$1\!/N$ expansion \cite{brmu}. Numer\-ical tests of our semi-classical 
relations for a variety of systems will be given in Ref.\ \cite{bkmr}. 

{\it Conclusions.--- } 
We have shown that quantum oscillations in spatial densities
can be derived without resorting to wave functions, but using 
the closed non-periodic orbits of the classical system. Our
one-dimensional result for $\delta\rho(x)$ is equivalent to that 
of \cite{ks}, but its derivation by the summation over classical orbits 
appears more transparent to us. We note that the semi-classical 
theory can be easily generalized to grand-canonical systems at 
finite temperatures \cite{temp}. Our results may become useful 
in the analysis of weakly interacting trapped fermionic gases (see, 
e.g., \cite{trapex}) for which the mean-field approximation is
appropriate. We present it as a challenge to verify the ``local 
virial theorem'' (\ref{taurho}) experimentally.

We are grateful to J. D. Urbina for helpful comments and to
A. Koch for numerical data used in the figure.


\end{document}